\documentclass{article}

\usepackage{arxiv}

\usepackage[utf8]{inputenc} 
\usepackage[T1]{fontenc}    
\usepackage{hyperref}       
\usepackage{url}            
\usepackage{booktabs}       
\usepackage{amsfonts}       
\usepackage{nicefrac}       
\usepackage{microtype}      
\usepackage{lipsum}		
\usepackage{graphicx}
\usepackage{natbib}
\usepackage{doi}
\usepackage{xcolor}

\usepackage{soul}

\newcommand{\hlparticipant}[1]{{\sethlcolor{lightgray}\hl{#1}}}

\usepackage{dirtytalk}

\usepackage{multirow}

\usepackage{ltablex}

\title{Right the docs: Characterising voice dataset documentation practices used in machine learning}

\date{March 20, 2023}	

\author{ \href{https://orcid.org/0000-0001-9205-0942}{\includegraphics[scale=0.06]{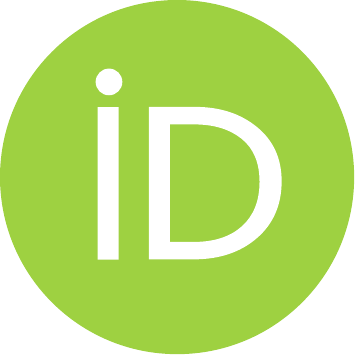}\hspace{1mm}Kathy Reid} \\
	School of Cybernetics\\
    Australian National University\\
    Canberra, Australia\\
	\texttt{kathy.reid@anu.edu.au} \\
\And
	\href{https://orcid.org/0000-0002-7895-458X}{\includegraphics[scale=0.06]{orcid.pdf}\hspace{1mm}Elizabeth T.~Williams} \\
	School of Cybernetics\\
    Australian National University\\
    Canberra, Australia\\
	\texttt{elizabeth.williams@anu.edu.au} \\
}

\renewcommand{\headeright}{}
\renewcommand{\shorttitle}{Right the docs: Characterising voice dataset documentation practices}

\hypersetup{
pdftitle={Right the docs: Characterising voice dataset documentation practices},
pdfsubject={cs.CL},
pdfauthor={Kathy Reid, Elizabeth T.~Williams},
pdfkeywords={dataset documentation, voice datasets, machine learning, documentation practices},
}

\begin{document}
\maketitle

\begin{abstract}

Voice-enabled technology, such as virtual assistants and smart speakers, are quickly becoming ubiquitous. They are constituted from machine learning (ML)-enabled components such as speech recognition and voice activity detection. However, these systems don't yet work well for everyone. They exhibit bias - defined here as systematic and unfair discrimination against individuals or cohorts of individuals in favour of others~\citep{friedman1996bias} - across axes such as age, gender and accent. 

ML is reliant on large datasets for training. Dataset documentation - such as metadata or descriptive information - is designed to give ML Practitioners (MLPs) a better understanding of a dataset's contents and characteristics. This is one approach MLPs use for preventing various forms of bias. However, there is a lack of empirical research on {\itshape voice} - spoken language - dataset documentation specifically. Additionally, while MLPs are frequent participants in fairness and bias research, little work focuses on those who work with {\itshape voice} data. Our work makes an empirical contribution to this gap. 

Here, we combine two methods to form an exploratory study. First, we undertake 13 semi-structured interviews with voice and adjacent MLPs, exploring multiple perspectives of voice dataset documentation practice. Using open and axial coding methods, we explore MLPs' practices through the lenses of roles and tradeoffs. Complementing and drawing from this work, we then purposively sample voice dataset documents (VDDs) for nine voice datasets. Our findings then triangulate these two methods, using the lenses of MLP roles and trade-offs. We find that current VDD practices are inchoate, inadequate and incommensurate. The characteristics of voice datasets are codified in fragmented, disjoint ways that often do not meet the needs of MLPs. Moreover, they cannot be readily compared or contrasted, presenting a barrier to practitioners' bias reduction efforts. 

We then discuss the implications of these findings for bias practices in voice data and speech technologies. Moving from exploration to action, we conclude by setting out a program of future work to address these findings --- that is, how we may ``right the docs''.
\end{abstract}


\section{Introduction, motivation and previous work} \label{IntroductionMotivation}
Voice-enabled technologies, such as virtual assistants and smart speakers, are ``going to scale'' through axes such as volume~\citep{kinsellaSmartSpeakerConsumer2020,bradleyBraceYourselfExplosion2020a,vandermeulenGartnerSaysWorldwide2016}, geographies~\citep{popovic2015voice,jones2020macsen,kendall2020understanding}, miniaturisation~\citep{bouraouiHardwareArchitecturesEmbedded2017}, expanding use cases~\citep{dale2020voice,brewer2022empirical,jesus2019voice} and use in multiple modalities~\citep{baevskiData2vecGeneralFramework2022}. Speech technology has become part of the fabric of modern {\itshape information infrastructures} --- the technical capabilities, social norms, organisational practices and economic mechanisms~\citep{bowker2009toward,turow2021voice} --- that collectively allow us to speak with machines and have them do our bidding. As the ubiquity of speech technology grows, so too does the potential societal impact of its bias. A person's poor voice interaction experience is no longer confined to a virtual assistant in the home, or to a mobile phone, but extends to the workplace, the car, healthcare, and customer service settings, {\itshape inter alia}. 

These systems employ machine learning (ML)-enabled components like automatic speech recognition (ASR) and voice activity detection (VAD). However, these systems don't yet work well for everyone~\citep{liu2022towards,ngueajio2022hey,feng2021quantifying}. They exhibit bias - defined here as systematic and unfair discrimination against individuals or cohorts of individuals in favour of others~\citep{friedman1996bias} --- across axes such as age, gender~\citep{tatman2017gender,tatman2017effects,garnerinGenderRepresentationOpen2020a}, race~\citep {koeneckeRacialDisparitiesAutomated2020}, nationality~\citep{hutiri2022bias}, and accent~\citep{hinsvark2021accented}. 

Identifying ways to mitigate bias here is challenging. Speech technologies are deployed in myriad spoken language contexts, and used by diverse cohorts of people for many tasks. Here, we choose to focus on two factors relevant to bias: dataset documentation and the practitioners who create and consume it. 

\subsection{Dataset documentation and its use by MLPs} \label{LiteratureReview}

The ML-enabled components in voice-enabled technologies require large datasets to be effective. {\itshape Dataset documentation} --- descriptive information characterising the nature, contents and provenance of a dataset --- affords MLPs a clearer understanding of a dataset's characteristics, allowing detection and prevention of some forms of bias. Similarly, {\itshape model documentation} --- descriptive information characterising the performance of a trained ML model against evaluative criteria --- affords MLPs the opportunity to detect and remediate some forms of bias. For these reasons, dataset and model documentation is well established in the extant literature as a bias detection and prevention tool. 

Firstly, we distinguish between datasheets and model cards. Datasheets or data statements describe a dataset which may be used alone or in conjunction with others for training an ML model. The focus here is on understanding the inputs to an ML model. Model cards, in contrast, characterise an ML model trained {\itshape from} a dataset; the focus here is on understanding the model's performance output. 

Foundational work in dataset documentation in the field of (primarily written) natural language processing (NLP) was done by Bender and Friedman~\citep{benderDataStatementsNatural2018}, who introduced the practice of {\itshape data statements}. Positioning data statements as a mechanism for addressing bias in the creation of language technologies, they proposed capturing information such as speaker demographics, annotator demographics, and the domain and context of the speech and text captured. Gebru et al~\citep{gebru2021datasheets} brings data provenance to the forefront of broader ML practice by outlining key areas a practitioner should consider, such as the purpose and intended use of the dataset, the objects it stores, how they're represented, the relationships between them, sources of error and noise, sensitivity and identification considerations, how the data was collected and labelled, and how the datasets are distributed and maintained. Boyd~\citep{boyd2021datasheets} seeks to empirically validate the utility of datasheets, and demonstrates their benefit by having MLPs ethically reflect on problematic datasets --- directly connecting datasheets as an artifact with improved practice. From the field of computer vision, Miceli et al~\citep{10.1145/3442188.3445880} also focus on praxis, emphasising the need for practitioner reflexivity in the production of ML datasets. Denton et al~\citep{denton2021genealogy}, also writing from a computer vision perspective, articulate the values and assumptions embedded by practitioners in the ImageNet dataset which is frequently used as an evaluative benchmark. 

In Costa-juss{\`a} et al~\citep{costa2020mt}, we see the adaptation of data statements and datasheets for datasets from NLP to other written language technologies --- in this case --- machine translation. Bandy and Vincent~\citep{bandy2021addressing} tie dataset documentation to the concept of technical debt, and retrospectively produce a datasheet for a text corpus. Pushkarna et al~\citep{pushkarna2022data}, based on their work with text corpora at Google, then introduce the concept of {\itshape data cards}, concentrating on descriptive information that cannot be inferred from the dataset itself. 



Concomitantly, there has been increasing attention on model documentation, primarily in the form of {\itshape model cards}, introduced by Mitchell et al in~\citep{mitchellModelCardsModel2019a}, and built on by Shen et al~\citep{shen2022model}, who produced a practitioner toolkit to aid in generic model card development. Crisan et al~\citep{crisanInteractiveModelCards2022}, recognising that many laypeople also use model documentation, develop an interactive approach to aid in model exploration. In~\citep{mcmillan2021reusable}, Mcmillan-Major et al seek to join both datasheets and model cards, proposing a standard format for datasets in NLP. 

However, data and model documentation in itself is not sufficient for tackling bias. A practitioner creates or consumes that documentation, providing a {\itshape feedback loop} which motivates practitioner {\itshape action}: re-balancing a training set, gathering more diverse data, or fine-tuning a model. 

Accordingly, recent work from Microsoft Research shifts the focus of inquiry to practitioners' use of dataset and model documentation~\citep{heger2022understanding} and approaches to fairness more broadly~\citep{holstein2019improving}. Heger et al find that dataset documentation practices are ``largely ad-hoc and myopic in nature'', with many practitioner needs unaddressed. Similarly, Holstein et al find, in a set of interviews with MLPs in industry, that while they saw the datasets as ``the most important place to intervene to improve fairness in their products'', the teams did not have in place processes - such as dataset documentation - ``to help support the collection and curation of balanced or representative datasets''. Work by Ashktorab et al~\citep{ashktorabFairnessEvaluationText2023}, from IBM, investigates how MLPs assess individual and group fairness in {\itshape text} classification models, providing recommendations for additional tooling. 

\subsection {The research gap} \label {ResearchGap}
People are increasingly using speech to interface with services and sources of support in the real world. ML-enabled voice systems continue to have pronounced biases; they work better for some people than others. If we wish to make the socio-technical systems of our world fairer, then we need to generate effective approaches for tackling bias in these systems. The approaches, motivations and actions of MLPs around dataset documentation have been shown to assist in this regard. However, there is a dearth of research here covering {\itshape spoken} language data --- the kind of data used to build voice systems. To address this gap, we pose the following research provocation: How might the voice dataset documentation practices currently in use by MLPs who work with voice datasets be characterised?


\section {Methodology} \label{Methodology}

To address these questions, we devise an exploratory study that combines two methods, one focusing on {\itshape ML practitioners} and the other on {\itshape dataset documentation artefacts}.  

Firstly, we focus on the people who create and/or consume dataset documentation as part of their practice. We undertake 13 semi-structured interviews with MLPs who work with voice or closely adjacent data. We explore their voice dataset document (VDD) approaches across the data and ML lifecycle, describe the challenges they face, and examine the trade-offs they must make. Secondly, we turn our attention to existing VDDs. VDDs represent how MLPs generate datasets and release them to the world --- they encode practices, beliefs and assumptions~\citep{birhane2022values}. We purposively select nine VDDs for their varied purposes, collection methods and source data. Drawing both from our literature review in Section ~\ref{LiteratureReview} and from the findings of the semi-structured interviews, we analyse the selected VDDs across seven categories. We compare and contrast our characterisations here to those identified through the interviews, highlighting unanticipated insights. 

Triangulating these two approaches affords a richer description of VDD practice than individual methods: the interviews contextualise the circumstances in which dataset documentation is produced and consumed and helps explain why VDDs have manifested in particular ways. Drawing from both methods also provides a stronger evidence basis for situating and directing further research work aimed at addressing bias in speech technologies. 

\subsection{Semi-structured interviews} \label{MethodologyInterviews}

Semi-structured interviews are established in the literature as an appropriate exploratory method for inquiring about phenomena, particularly with a view to setting out further actions. Moreover, they have been widely applied to the ML practice field~\citep{baier2019challenges,johnk2021ready,folstad2018makes}: this method is therefore an appropriate to apply here.

\subsubsection{Identification, selection and management of participants} \label{SelectionParticipants}

Firstly, institutional ethics approval was obtained. Potential participants were identified using professional networking tools, via snowball sampling, and via open source voice and speech technology projects. Our key inclusion criteria were (i) that the participant must work with voice or closely adjacent (e.g. acoustic or other signal) data, and (ii) be currently practicing in industry, academia or open source fields. Selection of participants was then done via purposive sampling to ensure representation of perspectives from diverse genders, disciplinary backgrounds, professional expertise and geographic locations, and to help establish trustworthiness of findings~\citep{campbell2020purposive,lincoln1985naturalistic,groves2011survey,ezzyCodingDataInterpreting2013}. 32 participants were approached, 19 declined, and 13 agreed to an interview. A summary of participants by characteristic is shown in Table ~\ref{table:ParticipantOverview}.

\begin{table*}[h]
\caption{Interview participant summary (n=13)}
\small
\centering
\begin{tabular}{ m{5cm} m{6cm} m{3cm}} 
\toprule
\multicolumn{2}{l}{Characteristic}               & Total        \\
\midrule
\multirow{2}{*}{Gender}                                                          & Female    & 3   \\
\cmidrule{2-3}
                                                                                 & Male      & 10      \\
\cmidrule{2-3}
                                                                                 & Other gender expressions      & 0      \\

\midrule

\multirow{4}{*}{Occupational field}                                              & Research scientist or academic   & 5   \\
\cmidrule{2-3}
                                                                                 & ML or NLP Engineer  & 2   \\
\cmidrule{2-3}
                                                                                 & Software Engineer  & 2   \\
\cmidrule{2-3}
                                                                                 & Data annotator   & 1   \\
\cmidrule{2-3}
                                                                                 & Developer Relations Advocate   & 2   \\
\cmidrule{2-3}
                                                                                 & UX Designer / researcher   & 1   \\
\midrule
\multirow{3}{*}{Country of residence}                                            & United States & 5         \\
\cmidrule{2-3}
                                                                                 & Australia & 3        \\
\cmidrule{2-3}
                                                                                 & South Africa  & 1         \\
\cmidrule{2-3}
                                                                                 & Aotearoa New Zealand  & 1         \\
\cmidrule{2-3}
                                                                                 & Nigeria  & 1         \\
\cmidrule{2-3}
                                                                                 & France  & 1         \\
\cmidrule{2-3}
                                                                                 & Canada  & 1         \\                                                                                 
\bottomrule
\end{tabular}
\label{table:ParticipantOverview}
\end{table*}

To track potential and confirmed participants through the interview pipeline, a custom {\verb|MySQL|} database was designed and hosted locally~\footnote{\href{https://bit.ly/AIES23-nocodb-database}{https://bit.ly/AIES23-nocodb-database}}. Interviews were conducted via the Zoom\texttrademark{} videoconferencing platform and recorded locally. Transcriptions were undertaken using the Rev\texttrademark{} cloud service. Corrections were done manually, and information that would personally identify participants was redacted. Participants were provided with a copy of the edited transcript, and were able to make further corrections and/or redactions. We concluded our interviews at 13 participants, as themes were becoming repetitive, and we had sufficient data to inform next research steps.

\subsubsection{Semi-structured interview design} \label{InterviewDesign}
We adopted an inductive approach, seeking to accumulate many perspectives around how VDDs are produced and consumed, whilst varying their contexts, applications and geographic sites of practice~\citep{creswell2003research}. Drawing from both Spradley and Minichiello et al ~\citep{spradleyEthnographicInterview1979, minichielloIndepthInterviewingResearching1990}, we structured our interview questions around ``the lifecycle of creating a voice dataset'' --- a ``grand tour'' approach, then dove deeper into topics related to bias as they arose --- a ``mini tour''. 

\subsubsection{Coding approach} \label{CodingApproach} 

 We performed an environmental scan of available qualitative analysis tools, and chose MaxQDA as it best aligned with our text-based needs and envisaged coding workflow. Based on our initial examination of the literature, previous MLP interviews, and those suggested in~\citep{saldana2021coding}, we identified several {\itshape a priori} categories. We then undertook coding of the 13 interviews, ascribing material to the {\itshape a priori} categories. We combined this with experimental open coding approach~\citep{williams2019art} to capture new categories as they emerged. This divergent step allowed us to gain an appreciation of the breadth of themes surfaced in the interviews. 
 
 We then performed axial coding to highlight the contextual conditions of VDD --- the circumstances of its production or absence, how it is used, and its consequences~\citep{ezzyCodingDataInterpreting2013}. We then performed selective coding per~\citep{corbinGroundedTheoryResearch1990}, allowing us to collapse and combine several codes into core categories for for analysis. This yielded 14 broad categorisations across a total of 1889 codes. Here, we focus on only three of those broad categories; the ML and data lifecycle, tradeoffs MLPs make, and challenges they face.

 Quotes from interview material here are pseudonymous.


\subsection {Document analysis of voice dataset documentation (VDD)} \label {CurrentDocumentation}

As a complementary method to the qualitative data produced from our semi-structured interviews, we then used document analysis --- ``a systematic procedure for reviewing or evaluating documents''~\citep{bowen2009document}.

\subsubsection {Identifying and selecting dataset documents for analysis} \label{IdentifyingSelectingDatasets} 

 Datasets used for ML are often released with accompanying documentation in the form of a dedicated web site, code repository or online catalogue entry. Additionally, some datasets contain a metadata or other descriptive file {\itshape within} the dataset. We considered all of these in scope for analysis, and correctly assumed that the specific form of documentation would vary between datasets. 

 To identify VDDs for analysis, we performed a web search~\footnote{A web search was used, as datasets are not as well indexed in scholarly databases.}, using the terms ``voice dataset'' or ``speech dataset'', and identified several. From those, we identified nine that differed in purpose, origin, compilation and context, providing a purposive sample. Specifically, we sampled datasets that varied by intended task; by whether the speech was elicited or spontaneous; the domain of speech, the curation rationale, funding source; license; and vocabulary size. A summary is provided in Table ~\ref{table:DatasetOverview}.

\begin{table*}[ht]
    \caption{Summary of voice dataset documents analysed}
    \tiny 
    \begin{tabular}{rlllllllll}
    \toprule

    \parbox[t]{1.80cm}{\raggedleft\textbf{Characteristic of the dataset or voice dataset document (VDD) }}
        & \parbox[b]{0.85cm}{\raggedright{Mozilla Common Voice~\citep{ardilaCommonVoiceMassivelyMultilingual2019}}}
        & \parbox[b]{0.85cm}{\raggedright{Librispeech~\citep{panayotovLibrispeechAsrCorpus2015}}}
        & \parbox[b]{0.85cm}{\raggedright{African languages in the field~\citep{gauthier2016collect}}}
        & \parbox[b]{0.85cm}{\raggedright{Voxceleb~\citep{chung2018voxceleb2,nagrani2020voxceleb,nagrani2017voxceleb}}}
        & \parbox[b]{0.85cm}{\raggedright{LDC 2000 HUB5 English Evaluation Speech}}
        & \parbox[b]{0.85cm}{\raggedright{TED-LIUM corpus~\citep{rousseauTEDLIUMAutomaticSpeech2012,hernandezTEDLIUMTwiceMuch2018}}}
        & \parbox[b]{0.85cm}{\raggedright{Free spoken digit dataset~\citep{zohar_jackson_2018_1342401}}}
        & \parbox[b]{0.85cm}{\raggedright{CHIME 5 Speech separation challenge dataset~\citep{barker2018fifth}}}
        & \parbox[b]{0.85cm}{\raggedright{LJSpeech Speech dataset~\citep{itoLJSpeech2017}}} \\

        \midrule 

        \parbox[t]{1.80cm}{\raggedleft{Type of document(s) analysed}}& \parbox[t]{1.20cm}{\raggedright{\href{https://commonvoice.mozilla.org}{CommonVoice website}, \href{https://github.com/common-voice/cv-dataset/}{GitHub repository}, related paper}} & \parbox[t]{1.20cm}{\raggedright{\href{https://openslr.org/12/}{Entry on OpenSLR website}, related paper}} & \parbox[t]{1.20cm}{\raggedright{\href{https://www.openslr.org/25/}{Entry on OpenSLR website, \texttt{README} file in dataset}}, related paper} & \parbox[t]{1.20cm}{\raggedright{\href{https://www.robots.ox.ac.uk/~vgg/data/voxceleb/}{VoxCeleb website}}, \href{https://web.archive.org/web/20200629153946/http://www.robots.ox.ac.uk/~vgg/data/voxceleb/meta/vox2_meta.csv}{Metadata file archived on archive.org}, related papers} & \parbox[t]{1.20cm}{\href{https://catalog.ldc.upenn.edu/LDC2002S09}{LDC Catalogue entry}} & \parbox[t]{1.20cm}{\raggedright{\href{https://lium.univ-lemans.fr/en/ted-lium2/}{TED-LIUM website}, \texttt{README} file in dataset}, related paper} & \parbox[t]{1.20cm}{\raggedright{\href{https://github.com/Jakobovski/free-spoken-digit-dataset}{GitHub repository}, \href{https://zenodo.org/record/1342401}{Zenodo dataset record}, \texttt{metadata.py} file in dataset }} & \parbox[t]{1.20cm}{\raggedright{\href{https://spandh.dcs.shef.ac.uk/chime_challenge/CHiME5/data.html}{Data page on CHIME website}}, \texttt{JSON file in dataset}} & \parbox[t]{1.20cm}{\raggedright{\href{https://keithito.com/LJ-Speech-Dataset/}{LJ Speech website}}}\\

        \midrule 

        \parbox[t]{1.80cm}{\raggedleft{Year of dataset initial release and latest version}}& \parbox[t]{1.20cm}{\raggedright{2018; 2023 (version 13)}} & \parbox[t]{1.20cm}{\raggedright{2015; no newer version}} & \parbox[t]{1.20cm}{\raggedright{2005; no newer version}} & \parbox[t]{1.20cm}{\raggedright{2017; 2018 (version 2)}} & \parbox[t]{1.20cm}{\raggedright{2005; no newer version}} & \parbox[t]{1.20cm}{\raggedright{2012; 2018 (version 3)}} & \parbox[t]{1.20cm}{\raggedright{2018; no newer version}} & \parbox[t]{1.20cm}{\raggedright{2018; CHIME 5 is the fifth iteration of a challenge}} & \parbox[t]{1.20cm}{\raggedright{2017; 2017 (version 1.1)}}\\ 

        \midrule 

        \parbox[t]{1.80cm}{\raggedleft{Intended language task}}& \parbox[t]{1.20cm}{\raggedright{Speech recognition}} & \parbox[t]{1.20cm}{\raggedright{Speech recognition, multilingual}} & \parbox[t]{1.20cm}{\raggedright{Speech recognition, monolingual}} & \parbox[t]{1.20cm}{\raggedright{Speaker identification, speech separation and cross-modal transfer, monolingual}} & \parbox[t]{1.20cm}{\raggedright{Speech recognition, monolingual}} & \parbox[t]{1.20cm}{\raggedright{Speech recognition, monolingual}} & \parbox[t]{1.20cm}{\raggedright{Speech recognition, monolingual}} & \parbox[t]{1.20cm}{\raggedright{Speech separation, monolingual}} & \parbox[t]{1.20cm}{\raggedright{Speech synthesis, monolingual}}\\ 

        \midrule 

        \parbox[t]{1.80cm}{\raggedleft{Nature of speech in dataset}}& \parbox[t]{1.20cm}{\raggedright{Elicited, large vocabulary, multiple domains}} & \parbox[t]{1.20cm}{\raggedright{Elicited, large vocabulary, out of copyright works}} & \parbox[t]{1.20cm}{\raggedright{Elicited, large vocabulary, multiple domains}} & \parbox[t]{1.20cm}{\raggedright{Spontaneous, large vocabulary, multiple domains}} & \parbox[t]{1.20cm}{\raggedright{Spontaneous, large vocabulary, multiple domains}} & \parbox[t]{1.20cm}{\raggedright{Spontaneous, large vocabulary, multiple domains}} & \parbox[t]{1.20cm}{\raggedright{Elicited, constrained vocabulary, spoken digits}} & \parbox[t]{1.20cm}{\raggedright{Spontaneous, large vocabulary, multiple domains}} & \parbox[t]{1.20cm}{\raggedright{Elicited, large vocabulary, non-fiction books publishes between 1884 and 1964}}\\

        \midrule 

        \parbox[t]{1.80cm}{\raggedleft{Motivation and funding source}}& \parbox[t]{1.20cm}{\raggedright{Ecosystem development; Grant-based for particular languages; additional funding from NVIDIA; otherwise funded by non-profit Mozilla Foundation.}} & \parbox[t]{1.20cm}{\raggedright{Research; funding unknown.}} & \parbox[t]{1.20cm}{\raggedright{Research; \href{https://web.archive.org/web/20220121164956/http://alffa.imag.fr/}{ALFFA Research Project}, funded by \href{https://anr.fr/}{agence nationale de la recherche}}.} & \parbox[t]{1.20cm}{\raggedright{Research by Oxford University, funded through {\href{http://www.robots.ox.ac.uk/~vgg/projects/seebibyte/}{EPSRC programme grant Seebibyte EP/M013774/1: Visual Search for the Era of Big Data.}}}} & \parbox[t]{1.20cm}{\raggedright{Commercial; Sponsored by National Institute of Standards and Technology.}} & \parbox[t]{1.20cm}{\raggedright{Research, funding not specified.}} & \parbox[t]{1.20cm}{\raggedright{Research, funding not specified.}} & \parbox[t]{1.20cm}{\raggedright{Research challenge sponsored by Google and Microsoft Research.}} & \parbox[t]{1.20cm}{\raggedright{Research, funding not specified, independent researcher.}}\\

        \midrule 

        \parbox[t]{1.80cm}{\raggedleft{Method of collection of dataset}}& \parbox[t]{1.20cm}{\raggedright{Volunteer speakers recorded on web-based platform.}} & \parbox[t]{1.20cm}{\raggedright{Secondary use dataset from Librivox volunteer audio book project~\citep{librivoxLibrivoxAcousticalLiberation2021}}, speaker consent obtained for audio book reading, but not for use in dataset.} & \parbox[t]{1.20cm}{\raggedright{Original dataset, volunteer speakers recorded in field.}} & \parbox[t]{1.20cm}{\raggedright{Secondary use dataset from YouTube; speakers' consent not provided.}} & \parbox[t]{1.20cm}{\raggedright{Original dataset, recruited speakers recorded via telephone.}} & \parbox[t]{1.20cm}{\raggedright{Secondary use dataset from TED videos; speaker consent unknown.}} & \parbox[t]{1.20cm}{\raggedright{Original dataset, speaker recruitment and recording unknown.}} & \parbox[t]{1.20cm}{\raggedright{Original dataset, speaker recruitment unknown, recorded in speakers' homes.}} & \parbox[t]{1.20cm}{\raggedright{Tertiary dataset, subset of Librispeech containing single speaker. Speaker consent not stated.}}\\
        
        \bottomrule
    
    \end{tabular}
    \label{table:DatasetOverview}
\end{table*}


\subsubsection {Identifying characteristics of dataset documentation to analyse}

Drawing from previous work on dataset documentation identified in Section ~\ref{IntroductionMotivation}, as well as broader reading in metadata and research infrastructure, we synthesised and adapted seven categories to account for {\itshape spoken} language. We ruled out of scope investigating more granular representations at this stage of the analysis, such as how language or accent were represented, and we cover this in separate work~\citep{EasyChair:9678}. 

In the category of {\itshape Dataset identification}, we define a persistent identifier as a uniquely identifying string, separate from the location of the dataset itself, which provides a referral to the current storage location of the dataset~\citep{zeng2016metadata} and version as a way to distinguish dataset releases over time~\citep{bhattacherjee2015principles}. 

In the category of {\itshape Intent, purpose and curation rationale}, we draw on the definition given by Schlangen; a language task is a mapping between an input and an output, and a dataset provides examples of this mapping~\citep{schlangen2020targeting}. Clear descriptions of intent and purpose are therefore important so the MLP can identify if the dataset is task-appropriate. We adapt ``curation rationale'' as given in Bender and Friedman~\citep{benderDataStatementsNatural2018} to {\itshape spoken} language, and define it as determining which speech utterances are included in the dataset, and why. 

For the category of {\itshape Dataset creation process, sources and actors}, we draw again from Bender and Friedman~\citep{benderDataStatementsNatural2018}, who place emphasis on understanding the social standpoint of annotators. For many speech tasks, written transcriptions are also required as inputs. Noting the work of Bucholz --- that transcription has both variation and politics in its production~\citep{bucholtz2007variation,bucholtz2000politics} --- we also identified whether the transcription method was provided. Referencing work on spoken language variation from corpus linguistics (e.g.~\citep {barbiers2007syntactic}), we also identified whether the source of elicited speech prompts was specified. 

Looking at {\itshape Characteristics of the dataset itself}, we drew from material on research data infrastructure. Working with ``big data'' presents many challenges to MLPs~\citep{kitchin2014data};  and so it is beneficial to provide an overview of the size, shape and constituency of the dataset. 

Triangulating our two methods, we then drew additionally from our exploratory interviews for the categories of {\itshape Constitution of the dataset by speaker, recording environment and spoken language attributes}. In our exploratory interviews ~\ref{Findings}, comprehending contents was a key consideration for many participants. Speech recognition requires a wide variety of voice samples, while speech synthesis needs many samples from a single speaker. It is therefore important that characteristics of the speech utterances captured in the data are clearly represented.  Knowing how voice data varies within a dataset helps the MLP to ``balance'' it: to ensure that data items with salient characteristics are included or excluded: 

\say{...Sometimes you really need to dig deeply into the corpus to find it. Sometimes you just don't find it. And sometimes this is well documented. ... This is important ... because we need to have a balanced corpus for training your system. And then also to be able to evaluate, gender wise, the performance of your system.}  \hlparticipant {Sand Brown}

Within these categories, we then drew both from the literature and from exploratory interviews to identify particular attributes to assess. Bender makes the case for clearly identifying the languages we work with in~\citep{bender2019benderrule}, and in Bender et al~\citep{benderDataStatementsNatural2018} advocates both for representing the languages in a dataset in {\verb|BCP-47|} format {\itshape and} providing a ``prose description'' of the language's ``axes of variation'': 

\say{So there's a lot of axes of variation where failure to consider it in the data curation becomes a big problem.} \hlparticipant {Tibetan Stone}. 

\hlparticipant {Sand Brown} further highlighted aspects of variance of spoken language to consider: 

\say{...But you want to ask them a few question about their mother's and father's native language, and a few question about their usage of the language, where do they live, and these kind of very preliminary sociolinguistic questions ....} \hlparticipant {Sand Brown}

Participant \hlparticipant {Tibetan Stone} highlighted additional areas of spoken language variance to scrutinise when evaluating trained models: 

\say{... We have a lot of folks who have code-switched data ... it's also domain variation or register variation, or all your training data is super formal ...} \hlparticipant {Tibetan Stone}. 

Code-switching is where the speaker alternates between two or more ``codes'' --- usually languages -- within a conversation ~\citep{auer2013code}. The domain of spoken language is usually taken to be the subject matter of the conversation, while register is how spoken language varies by social situation; we speak differently in formal and informal settings~\citep{fineganLanguageItsStructure2014}.

The category of {\itshape Models, benchmarks and academic papers} is adapted from Gebru et al~\citep{gebru2021datasheets}, who recommend documenting where a dataset has previously been used to guide MLPs in where it should {\itshape not} be used: \say {When you think about kind of building a dataset, it's easy to think about, \say{Okay, I'm building a dataset, it's going to be used for this. This is what I want it to be used for.} Unfortunately, people are going to use it for things you didn't intend.} \hlparticipant{Citrus Green}. Similarly, noting increasing calls for benchmarks to be tightly linked to the intended task of a dataset (e.g.~\citep{raji2021ai}), we included these in the analysis. 

For the category of {\itshape Privacy, bias, limitations and social impact} we draw again from Bender and Friedman, and Gebru et al~\citep{benderDataStatementsNatural2018,gebru2021datasheets}, who both underscore the importance of documenting privacy and sensitivity considerations of a dataset, and their potential social sequelae. For example, a dataset may reveal sensitive attributes of a speaker, such as their membership in a marginalised language group: consider a Uygur native speaker in Han-speaking China, or an Arabic native speaker in post-911 New York. We draw from previous bias work in speech technologies, previously cited in~\ref{IntroductionMotivation} and consider whether biases in the dataset are articulated, and any limits of application considered. 

\subsubsection{Performing the analysis}

To perform the analysis, we reviewed the documents for each dataset against the criteria in each category, and summarised our findings in Table~\ref{table:CurrentDatasetdocumentation}. If fulfilment of a criterion was implied but not explicit in the document(s), then we made a finding of ``Implied'' and provided a rationale. If a criterion was not applicable to a dataset, we made a finding of ``N/A'' --- for example, in speech synthesis datasets like ``LJSpeech'', speech samples are usually taken from only one speaker and so the number of unique speakers in the dataset is not applicable.

    \tiny
    \begin{tabularx}{\textwidth}{rlllllllll}
        \caption{Descriptions and data items included in current voice dataset documentation }
        \label{table:CurrentDatasetdocumentation}
    \\ \toprule
    \parbox[t]{2.95cm}{\raggedleft\textbf{Data item}}
        & \parbox[b]{0.85cm}{\raggedright\href{https://commonvoice.mozilla.org/en/datasets}{Mozilla Common Voice}}
        & \parbox[b]{0.85cm}{\raggedright\href{https://openslr.org/12/}{Librispeech}}
        & \parbox[b]{0.85cm}{\raggedright\href{https://www.openslr.org/25/}{African languages in the field}}
        & \parbox[b]{0.85cm}{\raggedright\href{https://www.robots.ox.ac.uk/\~vgg/data/voxceleb/}{Voxceleb}}
        & \parbox[b]{0.85cm}{\raggedright\href{https://catalog.ldc.upenn.edu/LDC2002S09}{LDC 2000 HUB5 English Evaluation Speech}}
        & \parbox[b]{0.85cm}{\raggedright\href{https://lium.univ-lemans.fr/en/ted-lium1/}{TED-LIUM corpus}}
        & \parbox[b]{0.85cm}{\raggedright\href{https://github.com/Jakobovski/free-spoken-digit-dataset}{Free spoken digit dataset}}
        & \parbox[b]{0.85cm}{\raggedright\href{http://spandh.dcs.shef.ac.uk/chime\_challenge/CHiME5/data.html}{CHIME 5 Speech separation challenge dataset}}
        & \parbox[b]{0.85cm}{\raggedright\href{https://keithito.com/LJ-Speech-Dataset/}{LJSpeech Speech dataset}} \\
    
        \midrule
        \endhead
        
        \multicolumn{10}{l}{\textbf{Dataset identification}} \\
        \parbox[t]{2.95cm}{\raggedleft{Persistent identifier for the dataset}} & No & No & No & No & \href{https://doi.org/10.35111/p7pz-x179}{Yes} & No & \href{https://zenodo.org/badge/latestdoi/61622039}{Yes} & \href{https://doi.org/10.21437/interspeech.2018-1768}{Yes} & No \\ 
        \parbox[t]{2.95cm}{\raggedleft{Dataset versioning}} & Yes & Yes & No & Yes & Yes & Yes & Yes & \parbox[t]{0.85cm}{\raggedright{Implied via yearly competition}} & Yes \\ 
        \parbox[t]{2.95cm}{\raggedleft{Dataset release date}} & Yes & \parbox[t]{0.85cm}{\raggedright{Implied through related paper}} & Yes & \parbox[t]{0.85cm}{\raggedright{Implied through related paper}} & Yes & \parbox[t]{0.85cm}{\raggedright{Implied through related paper}} & Yes & Yes & Yes \\ 
        
        \midrule
        \multicolumn{10}{l}{\textbf{Intent, purpose and curation rationale}}\\
        \parbox[t]{2.95cm}{\raggedleft{Executive summary or description}} & Yes & Yes & Yes & Yes & Yes & Yes & Yes & Yes & Yes \\ 
        \parbox[t]{2.95cm}{\raggedleft{Intended tasks or use cases}}& Yes & Yes & Yes & Yes & Yes & Yes & \parbox[t]{0.85cm}{\raggedright{Implied through GitHub repository tags}} & Yes & No\\ 
        \parbox[t]{2.95cm}{\raggedleft{Curation rationale}}& Yes & Yes & Yes & Yes & Yes & Yes & Yes & Yes & No\\ 

        \midrule
        \multicolumn{10}{l}{\textbf{Dataset creation process, sources and actors}}\\ 
        \parbox[t]{2.95cm}{\raggedleft{Dataset collection method}} & Yes & Yes & Yes & Yes & Yes & Yes & No & Yes & Yes \\ 
        \parbox[t]{2.95cm}{\raggedleft{For elicited speech, the source of prompts}} & \parbox[t]{0.85cm}{\raggedright{Implied through GitHub history}} & Yes & Yes & Yes & No & Yes & No & Yes & Yes \\ 
        \parbox[t]{2.95cm}{\raggedleft{For spontaneous speech, description of the annotation/transcription process}} & N/A & N/A & N/A & N/A & No & Yes & N/A & No & N/A \\ 
        \parbox[t]{2.95cm}{\raggedleft{For spontaneous speech, description of the annotators}} &N/A & N/A & N/A & No & N/A & No & N/A & N/A & N/A \\

        \midrule
        \multicolumn{10}{l}{\textbf{Characteristics of the dataset itself}}\\

        \parbox[t]{2.95cm}{\raggedleft{Structure of dataset, such as field mapping, described}} & Yes & Yes & Yes & Yes & Yes & Yes & Yes & Yes & Yes \\ 
        \parbox[t]{2.95cm}{\raggedleft{Dataset storage size provided}} & Yes & Yes & Yes & Yes & No & Yes & Yes & Yes & Yes \\  
        \parbox[t]{2.95cm}{\raggedleft{Overall hours of speech in dataset specified}} & Yes & Yes & Yes & Yes & Yes & Yes & Yes & Yes & Yes \\ 
        \parbox[t]{2.95cm}{\raggedleft{License specified}} & CC0 & CC-BY-4.0 & MIT & CC-BY-SA-4.0 & LDC User Agreement & CC-BY-NC-ND-3.0 & CC-BY-SA-4.0 & \parbox[t]{0.85cm}{\raggedright{Dataset specific}} & \parbox[t]{0.85cm}{\raggedright{Public domain}}  \\ 
        
        \parbox[t]{2.95cm}{\raggedleft{\# of distinct voices in dataset specified}} & Yes & Yes & Yes & Yes & \parbox[t]{0.85cm}{\raggedright{Implied via \# of conversations}} & \parbox[t]{0.85cm}{\raggedright{Yes, in paper}} & Yes & Yes & Yes \\ 
        
        \parbox[t]{2.95cm}{\raggedleft{\# of utterances in dataset specified}} & Yes & Yes & Yes & Yes & \parbox[t]{0.85cm}{\raggedright{No, only \# of conversations given}} & \parbox[t]{0.85cm}{\raggedright{Yes, in paper}}  & Yes & Yes & Yes \\ 
        
        \parbox[t]{2.95cm}{\raggedleft{Length of utterances given}} & Yes & No & \parbox[t]{0.85cm}{\raggedright{Yes, averaged}} & \parbox[t]{0.85cm}{\raggedright{Implied via each utterance having same length}} & No & No & \parbox[t]{0.85cm}{\raggedright{Implied via each utterance being a single digit}} & \parbox[t]{0.85cm}{\raggedright{Inferred via JSON file}} & \parbox[t]{0.85cm}{\raggedright{Yes, averaged}}\\ 
        
        \parbox[t]{2.95cm}{\raggedleft{Split information (test, train, dev etc) provided}} & Yes & Yes & \parbox[t]{0.85cm}{\raggedright{Yes, in data structure}} & No & No & \parbox[t]{0.85cm}{\raggedright{Yes, in data structure}} & Yes & Yes & \parbox[t]{0.85cm}{\raggedright{N/A, splits not used in speech synthesis}} \\ 
        
        \parbox[t]{2.95cm}{\raggedleft{Audio file type specified}} & \parbox[t]{0.85cm}{\raggedright{Yes, in data structure}}  & \parbox[t]{0.85cm}{\raggedright{Yes, in data structure}} & \parbox[t]{0.85cm}{\raggedright{Yes, in data structure}}  & No & \parbox[t]{0.85cm}{\raggedright{File type implied by sample file}} & Yes & Yes & Yes & Yes \\         
        \parbox[t]{2.95cm}{\raggedleft{Audio file format details (resolution etc) provided}} & No & Yes (some) & No & No & Yes & Yes & Yes & Yes & Yes \\ 
        
        \midrule
        \multicolumn{10}{l}{\textbf{How the dataset represents characteristics of the speaker(s)}}\\
        
        \parbox[t]{2.95cm}{\raggedleft{Representation or distribution of speaker accent}} & Yes & No & No & No & No & No & Yes & No & No \\ 
        
        \parbox[t]{2.95cm}{\raggedleft{Representation or distribution of speaker nationality}} & No & No & No & No & No & No & Yes & No & No \\ 
        
        \parbox[t]{2.95cm}{\raggedleft{Representation or distribution of speaker age}} & Yes & No & No & No & No & No & No & No & No\\ 
        
        \parbox[t]{2.95cm}{\raggedleft{Representation or distribution of speaker gender}} & Yes & Yes & \parbox[t]{0.85cm}{\raggedright{Not in dataset, but distribution specified in paper}} & Yes & No & \parbox[t]{0.85cm}{\raggedright{Yes, in paper}} & Yes  & Yes & \parbox[t]{0.85cm}{\raggedright{Implied, single speaker, gender specified}}  \tabularnewline 
        
        \parbox[t]{2.95cm}{\raggedleft{Representation of speaker occupation}} & No & No & \parbox[t]{0.85cm}{\raggedright{Not in dataset, but overview given in paper}} & No & No & No & No & No & No \\ 
        
        \parbox[t]{2.95cm}{\raggedleft{Representation of speaker language acquisition or heritage}} & No & No & No & No & No & No & No & No & No \\
        
        \parbox[t]{2.95cm}{\raggedleft{Representation of speaker educational attainment}} & No & No & No & No & No & No & No & No & No \\


        \\
        \multicolumn{10}{l}{\textbf{How the dataset represents the recording environment}}\\
        \parbox[t]{2.95cm}{\raggedleft{Constitution by recording hardware}} & No & No & No & No & \parbox[t]{0.85cm}{\raggedright{Implied (telephone}} & No & No & No & No \\ 
        \parbox[t]{2.95cm}{\raggedleft{Constitution by recording environment}} & No & No & No & \parbox[t]{0.85cm}{\raggedright{Implied (interview)}} & \parbox[t]{0.85cm}{\raggedright{Implied (phone conversations)}} & \parbox[t]{0.85cm}{\raggedright{Implied (TED talks)}} & No & Yes & No \\ 
        \\
        \multicolumn{10}{l}{\textbf{How the dataset represents characteristics of spoken language}}\\
        
        \parbox[t]{2.95cm}{\raggedleft{Dataset language(s) represented \\using a standard such as \texttt{BCP-47} or \texttt{ISO-639}}} & Yes & Yes & Yes & Yes & Yes & No & No & No & No\\ 
        
        \parbox[t]{2.95cm}{\raggedleft{Multilingual flag}} & \parbox[t]{0.85cm}{\raggedright{Implied through dataset structure}} & Yes & \parbox[t]{0.85cm}{\raggedright{Implied through dataset structure}} & No & No & No & No & No & No\\ 
        
        \parbox[t]{2.95cm}{\raggedleft{Representation of phonetic distribution or variation}} & No & No & No & No & No & No & No & No & No\\ 
        
        \parbox[t]{2.95cm}{\raggedleft{Representation of dialect, lexical or non-phonetic variation}} & No & No & No & No & No & No & No & No & No \\ 
        
        \parbox[t]{2.95cm}{\raggedleft{Representation of domain of speech}} & No & \parbox[t]{0.85cm}{\raggedright{Implied due to Librivox source}} & No & No & No & No & \parbox[t]{0.85cm}{\raggedright{Implied - digits}} & No & \parbox[t]{0.85cm}{\raggedright{Implied due to Librivox source}} \tabularnewline 

        \parbox[t]{2.95cm}{\raggedleft{Constitution by formality or register of spoken language}} & \parbox[t]{0.85cm}{\raggedright{Varies with prompt}} & \parbox[t]{0.85cm}{\raggedright{Varies with prompt}} & No & No & No & \parbox[t]{0.85cm}{\raggedright{Implied - TED talks}} & No & No & \parbox[t]{0.85cm}{\raggedright{Varies with prompt}} \tabularnewline 
        
        \parbox[t]{2.95cm}{\raggedleft{For spontaneous speech, \\whether code-switching is indicated}} & N/A & N/A & N/A & No & No & No & N/A & No & N/A \\

        \midrule
        \multicolumn{10}{l}{\textbf{Models, benchmarks and academic papers}}\\
        
        \parbox[t]{2.95cm}{\raggedleft{Benchmarks specified or linked to}} & \parbox[t]{0.85cm}{\raggedright{No, uses CER for eval'n but no benchmark}} & Yes, WSJ  & No & \parbox[t]{0.85cm}{\raggedright{Yes, previous speaker recog'n datasets}} & No & No & \parbox[t]{0.85cm}{\raggedright{No, uses WER and CER for eval'n but no benchmark}} & No & No \\ 
        
        \parbox[t]{2.95cm}{\raggedleft{Models trained from dataset specified or linked to}} & \parbox[t]{0.85cm}{\raggedright{Yes, specified in paper}} & \parbox[t]{0.85cm}{\raggedright{Yes, specified in paper}} & \parbox[t]{0.85cm}{\raggedright{Yes, specified in paper}} & \parbox[t]{0.85cm}{\raggedright{Yes, specified in paper}} & No & No & No & \parbox[t]{0.85cm}{\raggedright{Yes, in \href{https://spandh.dcs.shef.ac.uk/chime_challenge/CHiME5/results.html}{results page}}} & No \\ 
        
        \parbox[t]{2.95cm}{\raggedleft{Papers based on dataset specified or linked to}} & \parbox[t]{0.85cm}{\raggedright{Yes, on website}} & \parbox[t]{0.85cm}{\raggedright{Yes, on website}} & \parbox[t]{0.85cm}{\raggedright{Yes, on website}} & \parbox[t]{0.85cm}{\raggedright{Yes, on website}} & No & \parbox[t]{0.85cm}{\raggedright{Yes, on website}} & Yes & \parbox[t]{0.85cm}{\raggedright{Yes, in \href{https://spandh.dcs.shef.ac.uk/chime_challenge/CHiME5/results.html}{results page}}} & No \\ 
  
        \midrule
        \multicolumn{10}{l}{\textbf{Privacy, limitations and social impact}}\\
        \parbox[t]{2.95cm}{\raggedleft{Privacy or sensitivity statement of the dataset}} & \parbox[t]{0.85cm}{\raggedright{Some info on~\href{https://commonvoice.mozilla.org/en/about}{website}}} & No & No & \parbox[t]{0.85cm}{\raggedright{Has a ~\href{https://www.robots.ox.ac.uk/~vgg/terms/url-lists-privacy-notice.html}{privacy statement}}} & No & No & No & No & No  \\ 
        
        \parbox[t]{2.95cm}{\raggedleft{Social impact statement of the dataset}} & \parbox[t]{0.85cm}{\raggedright{Some info on~\href{https://commonvoice.mozilla.org/en/about}{website}}} &  No & No & No & No & No & No & No & No  \\ 
        
        \parbox[t]{2.95cm}{\raggedleft{Statement of biases in dataset}} & No & No & No & No & No & No & No & No & No \\ 
        
        \parbox[t]{2.95cm}{\raggedleft{Statement of limitations of dataset}} & No & No & No & No & No & No & No & No & No  \\ 
        \bottomrule 
    \end{tabularx}
    \normalsize

\section {Findings} \label{Findings}

Here, we triangulate our two methods. We characterise VDD practices through the two lenses of practitioner roles and trade-offs the practitioner must make; this layered approach helps provide a richer characterisation of VDD practices. Our interview data profferred other analytical angles such as the challenges faced by practitioners, or practices across the data and ML lifecycle. We are limited for space here, and believe that these two lenses provide a representative characterisation.  

\subsection {Characterising practices by role} \label{FindingsRole}

\subsubsection {Chefs, Diners, Scavengers and Hoarders: voice datasets are discovered, created, commissioned or accumulated}  \label{FindingsScavengers} 

Here, we draw on our interview data to characterise the ways in which an MLP produces and/or consumes a dataset and related VDDs as one of four roles: Chefs, Diners, Scavengers and Hoarders. We then go on to outline the implications for each role arising from our documentation analysis. These depictions emerged primarily from interview questions early in the data and ML lifecycle exploring how the practitioner came into contact with datasets and VDDs. 

We characterise as {\itshape Chefs} those MLPs who are provided with a dataset specification against which to create a voice dataset: \say {... we would have a data collection spec, [with a] percentage of different accents or gender or whatever.} --- \hlparticipant {Blueberry Pie}. {\itshape Chefs} are mostly likely to be {\itshape producers} of VDDs. 

{\itshape Diners} form a complement to {\itshape Chefs}, being the MLPs who are in a position to order voice datasets from commercial companies. These companies offer both bespoke options - à la carte - as well as subscriptions to regular dataset updates - a grocery box. There are many such providers: \say {So there are many companies that offer services in terms of annotating data, transcribing data. There are many companies that collect some data and sell data.} --- \hlparticipant {Sodium Silver}. Alternatively, an MLP may be a {\itshape Scavenger} --- where they must discover freely available voice datasets to meet their needs due to cost constraints: \say {... us open source folks we're scavengers, right? ... The ordering options are there ... and I've looked at them and they want tens of thousands of dollars, for access. And I'm like, \say{"I don't have that."}} --- \hlparticipant{Poppy Surprise}.  

{\itshape Hoarders}, in contrast to Scavengers, Chefs and Diners, do not have a clear intent in mind for the voice data they accumulate; they store it for some future, unspecified purpose in the hope that it will be of use. Voice data accumulated this way is usually a byproduct of business operations: \say {We know that often companies, they have a plan to extract and collect as much data as possible before they even know what it's potentially useful for.} --- \hlparticipant{Peaceful Purple}. 

\subsubsection{The focus of VDD practices differs by role} \label{FocusVDDRole}

For the {\itshape Scavenger}, dataset documentation is important to their discovery efforts - and their ability to comprehend the contents of a dataset when found. Based on our document analysis, their needs are currently poorly served. While eight of the nine datasets represented speaker gender, only two represented accent, and only one represented speaker nationality or age. Speaker occupation, language heritage or education attainment were absent, save for an overview of speaker occupation in the {\itshape African languages} VDD. There was very little information provided on the recording environments used, and the only representation of variance of spoken language tended to be the way in which the dataset language(s) were specified --- with five of the nine VDDs representing language using a {\verb|BCP-47|} or {\verb|ISO-639|} code. 

{\itshape Chefs} may produce documentation as part of their creation efforts, in doing so, must make decisions about how to represent that data. With both an absence of agreed or {\itshape} de facto standards for documenting voice and speech data, as well as a proliferation of some standards for language representation ~\citep{wrightStandardsLanguageTranslation2019}, some participants faced challenges in determining {\itshape how} some data items should be reported: \say{There is no unified format. Everybody has their own JSON~\footnote{JSON is a data structure format commonly used for voice data} that might have similar information. So in a similar way, why isn't the W3C involved in this in the sense that give us some guidelines here?} --- \hlparticipant{Blueberry Pie}. 

Another {\itshape Chef} practitioner faced similar data representation dilemmas in regard to dialect, contending with the level of granularity required: \say{...what if we label what dialect they are speaking in? Or what if they self label what dialect they think they are speaking in? Then we do things like how about we review this? Meaning let's write whether we think this is pronounced correctly. It's either yes or no. Okay. Wait, what if we can label every single character in the sentence and say whether the character was pronounced correctly?} \hlparticipant{Emerald Green}

The lack of unified VDD format was also experienced while undertaking the documentation analysis in Table ~\ref{table:CurrentDatasetdocumentation}: attempting to ascertain whether a dataset had documented a criterion required painstaking cross-checking between websites, papers and \texttt{README} files. 

Both {\itshape Scavengers} and {\itshape Diners} need to know whether a voice dataset is useful for their intended purpose: \say{...the dataset documentation would give me an idea, does this dataset work for my application? ... Is this dataset going to be useful?} --- \hlparticipant{Citrus Green}. Drawing from our document analysis, it appears {\itshape Scavengers} and {\itshape Diners} are well served by current VDDs --- all nine datasets examined provided an executive summary or description, and eight of the nine provided both intended tasks or use cases, as well as a curation rationale. 

However, even if a dataset appears to meet an MLP's need based on the contents of the VDD, variation in how the dataset is transcribed can be problematic, requiring that the MLP spend time ``listening to the data'': \say{...All transcription is subjective. And so each of these databases will have been transcribed by different people, maybe following different conventions, and those conventions are especially important with semi words, ums and uhs and mm-mms, and stuff like that.} --- \hlparticipant{Blueberry Pie}. Another salient example here deals with the lack of variation of accents in the data not being apparent from the VDD, a realisation the practitioner makes only {\itshape after} listening to the data, and having to cross-check with the dataset's related academic paper: \say{...I had worked with it for a while, I thought I knew the data. It was a very popular dataset. And it wasn't until I started listening to it, that I realized that these are only North American voices. It wasn't obvious to me into that. And then I went back and I read the paper, the actual paper ... and it was explicit like, yes, they chose voices that were North American. And it's something simple as that, you don't know until you start listening to the data.} --- \hlparticipant{Citrus Delight}

While {\itshape Hoarders} may not know yet what tasks they wish to task of their datasets, VDDs are still relevant for this role. Hoarders may still wrestle with how to represent the data they are collecting. Here, there was a desire to create VDDs that allowed the broadest scope of future use for the data: \say{...it's always good to document, to label your data to the maximum extent that you can in terms of fidelity.} --- \hlparticipant{Raspberry Wine}. 

This desire links to the previously identified challenge of there being no standard approach to documenting spoken language data --- the desire to chronicle datasets with high fidelity places additional onus on the MLP to define how the data is represented. We see here a tendency to reproduce that which has come before: \say{...we didn't put a lot of thought into the choosing of the structure of the [Dataset] dataset, because we just used it as it was. And the reason that we chose the [Dataset] as an example dataset was because it was a fairly common, well-known speech recognition dataset}. This effect serves as a reinforcing loop, anchoring practice to the status quo. 

\subsection{Characterising practices through trade-offs the practitioner must make } \label{FindingsTradeOffs}

Changing our analytical lens, we now explore practices by exploring the tradeoffs a practitioner must make. Drawing from the field of social learning theory, Wenger-Trayner and Wenger-Trayner~\citep {wenger2014learning,wengerCommunitiesPracticeSocial2000} hold that practitioners operate across multiple disciplinary communities, in what they term as a ``landscape of practice''. Applying this conceptualisation to VDDs, a practitioner may need to span communities such as data engineering, machine learning, metadata specification and linguistics; each with their own rites, standards and accepted practices. These practices may be in tension, requiring the practitioner to make trade-offs. Exploring these helps us to more richly characterise VDD practice. While our interview data uncovered myriad trade-offs, we focus here on the most frequently recurring. 

\subsubsection{Needing big datasets, and also needing to store them} 

\say{The problem is, data gets big. And then you have a problem, right?} --- \hlparticipant{Auric Gold}

Speech technologies require hundreds to tens of thousands of hours of voice data, depending on the task \citep{reidMoreVoiceLess2021}. However, this volume presents numerous hurdles. A very practical one surfaced in our interviews was that faced by a {\itshape Chef} who created voice datasets, and needed to store them on a server. His frustration at having to frequently move datasets was palpable:  \say{Yeah. We would find somewhere on [University web server] we'd be, \say{Oh yeah. No, we'll serve it off our little file server here and it'll be no worries.} And we'd put it up there and we'd create a website for it. And we'd point people at the website. And then the IT guys would go, \say{Oh yeah, no. We don't want to do that  [...] We're going to shut that down. You're going to have to find somewhere else to put that.}} \hlparticipant{Raspberry Wine}. 

One mechanism that exists to overcome this limitation is the use of a {\itshape persistent identifier} - a data item we included in our document analysis in the category of {\itshape Dataset identification}. Only three of the nine datasets were found to have persistent identifiers applied, and these were verified using Crossref \footnote{\href{https://search.crossref.org/}{https://search.crossref.org/}}. On a more positive note, looking at the category of {\itshape Characteristics of the dataset itself}, all datasets bar one indicated storage requirements, and all provided the number of hours of overall speech in the dataset. 

\subsubsection{Needing big datasets, and needing to understand what's in them}

We also identified tradeoffs that the MLP had to make in comprehending the contents of a voice dataset. Earlier, in ~\ref{FocusVDDRole}, we showed how the MLP had to ``listen to the data'' when variation in speech samples was not well described by VDDs; that is, where the practitioner was compensating for the inadequacy of the VDDs by adopting an additional comprehension practice. The size of voice datasets makes this practice more onerous, as highlighted by an anecdote one participant related: \say{We ended up 12,000 recordings, which was humanly transcribed and those 12,000 recordings equated to 20 hours of speech. So we literally had a team of people listening to recordings and typing the recordings out verbatim.} --- \hlparticipant{Sodium Silver}. 

This again points to the need for more focus on capturing data related to recording environment in particular: \say{And with a hundred thousand hours of data, how are you going to listen to all that as one person especially? You can't. You can randomly sample and hope for the best that you catch something. But if you precisely knew exactly the conditions of the recordings and all that stuff, if you could control all that then I think you could do a much better job.} --- \hlparticipant{Poppy Surprise}. 

Triangulating this with our VDDs in the category of {\itshape How the dataset represents the recording environment}, we find that only the CHIME-5 dataset provided explicit information on the recording environment. This is likely due to its relevance in the dataset's purpose of speech separation. Other datasets implied some recording information - such as the HUB5 dataset being of recorded telephone conversations. Again, we find that VDDs are inadequate for practitioners' needs, and fragmented in how they present information.




\section {Limitations} \label{Limitations}

\paragraph{Dimensionality reduction} 
We note here that there are several layers of dimensionality reduction inherent in our interview collection and coding methodology. Although interviews were captured via video, only the audio was transcribed. This removed facial expressions and body language from the data under consideration. Further, the interviews were rendered from audio into text, again reducing dimensionality. 

\paragraph{Additional methods to triangulate findings} 
We recognise the small, although purposive sample of participants and datasets in our exploratory study. We now intend to administer a questionnaire to a broader group of MLPs, to allow triangulation with these initial findings, and to help validate the tentative conclusions for intervention we have identified here. 

\paragraph{Only publicly knowable datasets were analysed}
In identifying and selecting datasets for analysis, we recognise that our approach was limited to only publicly knowable datasets; private and/or proprietary datasets used internally by organisations may exhibit very different dataset documentation practices, although this is unlikely based on the work of~\citep{heger2022understanding, holstein2019improving}. 

\section {Implications and future work}

\paragraph{Towards a standard VDD for spoken language datasets} \label{FutureWorkStandards}
Our work here has shown that while there are inchoate attempts to create VDDs, their utility for MLPs is hampered by several drawbacks. Significantly, VDDs analysed here were not readily commensurable; they contained a patchwork of information in varying formats requiring significant effort to compare~\citep{busch2011standards}. Our work further highlighted the challenges MLPs face where they are also dataset documentation {\itshape producers} - such as the {\itshape Chef} role - in choosing representation formats for VDDs. 

A standard data statement format for spoken language datasets is likely to go some way to addressing these lacunae. We note similar efforts occurring on platforms that host ML datasets~\citep{ozeaniModelCards2022}, and acknowledge the challenges inherent in attempting to both formulate and drive uptake of standards. 

\paragraph{Tooling for constructing VDDs} \label{FutureWorkTools}

Even if a VDD format can be decided upon, tooling is still required to aid the MLP in creating VDDs. Many of the datasets we examined did not represent characteristics of the speaker; this data needs to be captured from the speaker during recording. Here, one interview participant suggests a data capture platform change to address this: \say {I would say that each time a new speaker is registered to the system, is going to start making a recording, we should have a nice interface, an easy to use interface, to quickly fill all the information that we need.} --- \hlparticipant{Sand Brown}

Additional tooling is also needed to appraise voice datasets for characteristics such as phonetic and lexical distribution. Such tools exist in isolation, but are not currently bundled for ease of use by MLPs. 

\paragraph {Data infrastructure literacy for voice MLPs} \label{FutureWorkDataLiteracy}
Gray et al~\citep{gray2018data} introduce the concept of ``data infrastructure literacy'', referring to both the data analytics competencies a practitioner must have, as well as their ability to influence broader socio-technical infrastructures in which their practice is situated. Our characterisation of current voice dataset documentation practices as inchoate invites further empirical work on {\itshape voice} data literacy appraisals.

\section {Conclusion} \label{Conclusion}

Here, we have situated voice dataset documentation (VDD) practices conducted by machine learning practitioners (MLPs) within broader efforts to reduce bias in ML-enabled speech technologies as they go to scale. We first provided a brief literature review of ML-related dataset documentation work, identifying that spoken language dataset practices are under-studied. 

We presented an exploratory study that combined two methods --- semi-structured interviews and document analysis --- to provide a rich characterisation of practices surrounding VDDs. We find that current VDD practices are currently inchoate, inadequate and incommensurate. Drawing from these findings, we propose actions that seek to ``right the docs'' --- giving practitioners the tools they need to prevent and detect bias and help speech technologies work better for more people. 


\section{Acknowledgements}
Thanks are extended to Glen Berman for feedback on this paper. Kathy Reid’s PhD research is funded by an Australian Research Training Scholarship and via the Florence Violet MacKenzie Scholarship. Kathy Reid holds a Research Partnership with Mozilla Foundation however this Partnership does not provide research funding. Kathy extends her heartfelt gratitude to the interview participants who gave generously of their time, knowledge and experience, and to the School of Cybernetics PhD cohort for their moral and intellectual support.

\bibliographystyle{unsrtnat}
\bibliography{datasetdoc-voice-exploratory} 




\end{document}